# Partout: A Distributed Engine for Efficient RDF Processing

Luis Galárraga
Max-Planck Institute for Informatics
Saarbrücken, Germany
lgallara@mpi-inf.mpg.de

Katja Hose
Department of Computer Science
Aalborg University, Denmark
khose@cs.aau.dk

Ralf Schenkel
Max-Planck Institute for Informatics
Saarbrücken, Germany
schenkel@mpi-inf.mpg.de

## ABSTRACT

The increasing interest in Semantic Web technologies has led not only to a rapid growth of semantic data on the Web but also to an increasing number of backend applications with already more than a trillion triples in some cases. Confronted with such huge amounts of data and the future growth, existing state-of-the-art systems for storing RDF and processing SPARQL queries are no longer sufficient. In this paper, we introduce PARTOUT, a distributed engine for efficient RDF processing in a cluster of machines. We propose an effective approach for fragmenting RDF data sets based on a query log, allocating the fragments to nodes in a cluster, and finding the optimal configuration. PARTOUT can efficiently handle updates and its query optimizer produces efficient query execution plans for ad-hoc SPARQL queries. Our experiments show the superiority of our approach to state-of-the-art approaches for partitioning and distributed SPARQL query processing.

## 1. INTRODUCTION

The increasing interest in Semantic Web technologies led to a rapid growth of available semantic data on the Web. Especially advances in information extraction [8, 19, 29, 33] enabled efficient and accurate extraction of knowledge from natural language text and its representation in a machine-readable format – RDF (Resource Description Framework). DBpedia[1], for instance, has now reached a size of 3.6 million entities and 1 billion RDF triples extracted from Wikipedia. As the number of Wikipedia articles increases every day and as information extraction techniques are still being improved, DBpedia and similar knowledge bases are likely to keep on growing. Some commercial data sets are even bigger by several orders of magnitude; according to the W3C, commercial data sets have already exceeded the 1 trillion triples barrier[2].

[1] http://dbpedia.org/
[2] http://www.w3.org/wiki/LargeTripleStores

It is not only the amount of data provided by a source that is growing but also the number of sources as the steady growth of the Linked Open Data (LOD) cloud[3] [2] has shown. LOD sources interlink their data by explicitly referencing data (URIs) provided by other sources and therefore building the foundations for answering queries over the data of multiple sources. Furthermore, more and more small RDF data sets without query processing interfaces become available on the Web. The data of such sources can usually be downloaded and processed locally. In the past few years, such data was crawled for the Billion Triple Challenge and resulted in about 1.5 billion triples in 2012[4].

Query processing in these scenarios is challenging because of the different ways in which sources can be accessed. Some sources provide SPARQL endpoints, others are available as downloadable data dumps, and still others as dereferenceable URIs[5]. This led to a variety of approaches for query processing that ranges from downloading the data during query processing [11, 13, 16] to the application of techniques known from distributed database systems [17, 26, 31] as a number of SPARQL endpoints, for instance, resembles a mediator-based or federated database system. The main disadvantage of these systems is the lack of control over the data, i.e., there is no guarantee on response time or that the data is available during query evaluation or that a source is answering a subquery at all. Furthermore, statistics that are essential for query optimization are hard to obtain and network communication delays are unpredictable.

An alternative approach to answer queries over multiple RDF sources corresponds to data warehousing, where the data is downloaded from the Web, collected in a huge triple store, and updated from time to time. Query processing in such a setup strongly benefits from efficient centralized query optimization and execution. Still, the ever-growing amount of RDF data will sooner or later result in scalability problems for a single machine. There are two widely used approaches to solve this problem: buying bigger machines (expensive mainframes) that can hold and process most of the data in main memory (centralized processing), or distributed query processing in a cluster of machines. As the first solution will sooner or later reach its limits in terms of available funding or scalability, applying a scale-out architecture and data partitioning with several cooperating machines based on commodity hardware is a good alternative, which has not yet been paid much attention to in this context. Although MapReduce can be used to process SPARQL queries, the immense overhead caused by starting the jobs makes it inferior to standard techniques for distributed query processing [14]. Similar to distributed databases and data warehouses in general, where data is often collected and stored to serve a particular use case, it is possible for many applications to derive a representative query workload because the data is often accessed in a similar way, e.g., simply because users access the data via a Web form generating SPARQL queries for a back-end triple store. By exploiting this information for partitioning the data, we obtain an additional gain in performance.

[3] http://linkeddata.org/
[4] http://challenge.semanticweb.org/
[5] An HTTP lookup of a URI provides a set of RDF triples with facts about the entity identified by the URI.

In this paper, we propose PARTOUT[6], a distributed engine based on a scale-out architecture for RDF data storage and distributed SPARQL query processing in a cluster of machines. One of the most important cost factors for query processing in such a setup is communication between nodes, which can be minimized by partitioning the data in a way that allows queries to be executed over a minimum number of nodes. We exploit a representative query load for partitioning and allocating the data to multiple hosts. Since access patterns of many applications are regular, such a query load is often available, or at least information about frequently co-ocurring subqueries can be collected. In contrast to the relational case where data comes in a natural partitioning into relations, RDF data corresponds to a single relation with three attributes so that queries involve a high number of self joins. Thus, standard distributed relational approaches are not directly applicable.

In summary, this paper makes the following contributions:

- PARTOUT[6], a system for scalable RDF storage that can efficiently handle updates to the data,
- a novel data partitioning and allocation algorithm for RDF data in consideration of a given sample query load that finds the optimal configuration of a cluster of machines, and
- an optimizer for distributed SPARQL query processing and a cost model for the proposed architecture.

Our evaluation shows that our approach is superior to state-of-the-art approaches for distributed SPARQL query processing.

The remainder of this paper is structured as follows. After having discussed related work and preliminaries in Sections 2 and 3, Section 4 presents a novel method for query load aware RDF partitioning and allocation. Section 5 then presents algorithms for efficient query processing and optimization in this setup and Section 6 evaluates the proposed algorithms and discusses the results. Section 7 finally concludes the paper with a summary and an outlook to future work.

## 2. RELATED WORK

We identify four main branches of related work: (i) centralized and (ii) decentralized approaches for RDF processing as well as (iii) parallel processing in clusters, and (iv) fragmentation and allocation.

---

[6]pronounced like the French word *partout*; the name is a combination of the terms *part*ition and scale-*out*

### Centralized Approaches

Building efficient databases for storing and querying RDF on a single machine has been a very active research topic for some years. Considering RDF data as a big table with three columns (subject, property, and object), the data can be stored in relational databases [28]. For RDF data, however, optimization techniques can be employed that are usually too expensive for standard relational databases, e.g., heavy indexing as used by RDF-3X [22] and Hexastore [35].

In addition, alternative storage formats have been proposed, such as property tables [18] combining multiple properties with the same subject and column stores [1, 32] splitting up triples into separate columns for subject, property, and object.

### Distributed Approaches

As more and more RDF knowledge bases become available on the Web, research also has to consider how to efficiently answer queries over multiple independent RDF sources that are accessible on the Web. Considering autonomous SPARQL endpoints, techniques known from distributed query processing in federated and mediator-based databases can be adapted [17, 26, 31]. Query processing over non-SPARQL endpoints can be realized by downloading the data during runtime [11, 13, 16] with or without the need for indexes or by applying the data warehouse approach, i.e., downloading all available data and relying on efficient centralized approaches.

Literature has also proposed approaches for RDF processing based on P2P systems [34]. By using an appropriate overlay design based on a hash function, these systems are able to assign RDF triples to different peers and find them efficiently during query processing. Focusing on individual triples, however, evaluating queries involving multiple joins can be very expensive in these systems.

### Parallel Processing in Clusters

When sources are organized in a cluster, query processing can make extensive use of parallel processing techniques, as applied by some commercial systems [3, 9]. Bigdata[7], for instance, uses B+ tree indexes to organize the data, dynamically partitions the B+ tree into index partitions (key-range shards), and assigns the partitions to the servers. The Virtuoso Cluster Edition[8] partitions the data using hashing and allows to configure which part of the data to use for this purpose, e.g., the first 5 characters of a string. Other systems, such as the OWLIM Replication Cluster[9], increase throughput by replicating the data to all nodes. However, none of these approaches considers query loads and therefore cannot improve performance by adapting to the way in which the data is actually used during query evaluation so that complex queries can become expensive.

As an alternative to applying techniques known from parallel databases, an increasingly large number of systems uses MapReduce [7]-style techniques to efficiently process RDF data in clusters [5, 15, 25]. A common disadvantage shared by all of them is the large response time of MapReduce processes. A number of proposals combine distributed storage and MapReduce, e.g., using HBase [25] or graph partition-ing [14]. Huang et al. [14] use a graph partitioner for splitting the RDF graph and partially replicate triples at the borders of partitions. When a query cannot be completely answered by the partitions alone, MapReduce is applied.

---

[7]http://www.bigdata.com/
[8]http://virtuoso.openlinksw.com/
[9]http://www.ontotext.com/owlim/replication-cluster

## Fragmentation and Allocation

In general, there are two important steps when setting up a distributed database system: fragmentation (partitioning the data into smaller fragments) and allocation (assigning the fragments to hosts). In the context of relational databases, techniques for horizontal (tuple-wise) and vertical (attribute-wise) partitioning according to a given query load have been published already about twenty years ago. Today, these techniques are employed by commercial systems in the context of relational databases [20, 24, 27], they are deeply integrated into the system to efficiently make use of internal statistics, cost models, and the optimizer. The main problem when trying to apply these approaches to RDF data and SPARQL queries is that they were developed for scenarios with different characteristics, e.g., joins between multiple relations vs. many self-joins for each query in case of RDF data.

More recently, graph-partitioning techniques have been proposed to find an appropriate partitioning of relations. Schism [6], for instance, constructs a graph where tuples correspond to nodes and edges connect nodes when they are accessed in the same transaction. The usually very high number of triples referenced by a SPARQL query during its evaluation leads to a very limited applicability to large RDF data sets.

Without consideration of a query load, [14] applies state-of-the-art graph-partitioning techniques on the RDF data graph itself, i.e., nodes correspond to entities occurring at subject positions in the data and two nodes are connected if a triple exists that refers to both entities (subject and object). In order to increase the number of SPARQL queries that can be answered without cross-partition joins, the approach uses replication, i.e., triples involving nodes at the border of a partition are replicated to multiple partitions.

Further systems applying replication have been proposed on top of P2P architectures. Still, for our problem scenario we cannot make use of them because they either consider each triple separately [12] or make assumptions on queries [30] that do not conform to the general-purpose setup we are considering in this paper.

## 3. PRELIMINARIES

## RDF and SPARQL

The Resource Description Framework[10], short RDF, is a data format proposed by the W3C for representing information about subjects in a machine-readable format. Subjects are represented as URIs, and statements about subjects are made in form of triples of the form (subject, property,object), where property is a URI and object is either a URI or a string literal. Thus, URIs identify entities (e.g., http://dbpedia.org/page/Berners-Lee), types (e.g., ex:Person), properties (e.g., rdfs:label), or values (mailto:em@w3.org). Collections of RDF triples can be represented as graphs, with subjects and objects as nodes and an edge from s to o with label p whenever triple (s,p,o) exists.

[10] http://www.w3.org/TR/rdf-concepts/

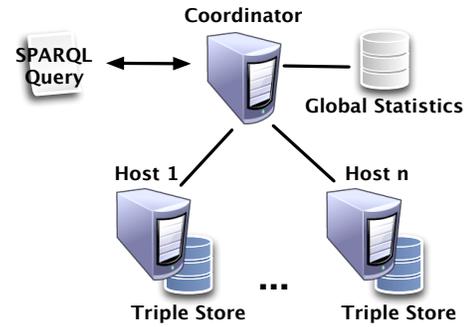

**Figure 1:** PARTOUT **System Model**

SPARQL[11] is a query language for RDF, proposed by the W3C. At its core are *basic graph patterns* that consist of one or more *triple patterns*, which are in turn triples where each of the subject, property, and object may be a variable, denoted by a leading *?*. Shared variable names among different triple patterns connect them to a graph pattern. SPARQL SELECT queries then retrieve bindings for (some of) the variables. Triple patterns can be marked as optional, and the value of variables can be constrained by additional FILTER expressions with operators such as $<, =,$ and $>$ for numerical values and string functions such as regular expressions. Multiple graph patterns can be combined through a *UNION* operator.

We formally represent a SPARQL query $q$ as its set of graph patterns $B(q) = \{b_1, \ldots, b_m\}$. Each graph pattern $b = (T(b), O(b), F(b))$ is represented by its (non-optional) triple patterns $T(b)$, its optional triple patterns $O(b)$, and its filter predicates $F(b)$.

## System Model

The general architecture of our system is depicted in Figure 1. It consists of a dedicated central coordinator and a cluster of $n$ hosts that store the actual data. As discussed in more detail in Section 5, the central coordinator is responsible for distributing the RDF data among the hosts, building an efficient distributed query plan for a SPARQL query, and initiating query execution. It utilizes global statistics of the RDF data for query planning. Each of the $n$ hosts runs a triple store, which in our implementation conforms to an adapted version of RDF-3X [21, 22].

## 4. LOAD-AWARE PARTITIONING

The partitioning and allocation process we propose has three main steps: (i) extract representative triple patterns from the query load applying normalization and anonymization (Section 4.1), (ii) use the extracted information to partition the data into fragments defined by fragmentation predicates (Section 4.2), and (iii) allocate the obtained fragments to hosts (Section 4.3). Table 1 shows an overview of our notation.

### 4.1 Extracting Relevant Information from a Query Load

We assume that we are given a query load $QL = \{q_1, \ldots, q_L\}$ of SPARQL queries, possibly including identi-cal queries. This can either be collected from a running system or estimated from queries in applications accessing the RDF data. For each such query $q \in QL$, we normalize and anonymize its (optional and non-optional) triple patterns in the following way: we replace infrequent URIs and literals in triple patterns with variables to avoid overfitting to the query-load, i.e., building fragments that contain only very few triples and that are accessed only by very few queries. Frequent URIs and literals, however, will not be pruned – the frequency threshold $\Theta$ above which URIs and literals are normalized is a parameter determined by the application scenario. We never normalize properties. Thus, for each triple pattern, we obtain its anonymized version $\omega(p)$, where each variable is replaced by the same anonymized symbol $\Omega$. We denote the set of all normalized and anonymized triple patterns included in the query load $QL$ by $\Phi(QL)$.

[11] http://www.w3.org/TR/rdf-sparql-query/

| | |
|---|---|
| $q$ | SPARQL query |
| $B(q)$ | graph patterns of $q$ |
| $T(b)$ | non-optional triple patterns of graph pattern $b$ |
| $O(b)$ | optional triple patterns of $b$ |
| $F(b)$ | filter predicates of $b$ |
| $QL$ | query load |
| $\Omega$ | anonymized variable |
| $\omega(p)$ | anonymized version of triple pattern $p$ |
| $\Theta$ | normalization threshold |
| $\Phi(QL)$ | set of anonymized normalized triple patterns in $QL$ |
| $G$ | a graph |
| $G(QL)$ | global query graph |
| $G(QL, M)$ | global fragment query graph |
| $M'$ | set of minterms for all simple predicates |
| $M$ | set of minterms for reduced set of simple predicates |
| $m$ | minterm and the corresponding fragment (long form: $T_m$) |
| $F$ | sets of fragments |
| $T$ | set of triples to fragment and allocate |
| $p, r$ | triple patterns, possibly normalized and anonymized |
| $f(p)$ | frequency of a triple pattern in $QL$ |
| $f(m)$ | frequency of a minterm (based on patterns in $QL$) |
| $s(m)$ | size of $m$ (in number of triples) |
| $n$ | number of hosts in the cluster |
| $h$ | a host in the cluster |
| $F_h$ | fragments assigned to host $h$ |
| $T_h$ | triples assigned to host $h$ |
| $L(m)$ | load induced by fragment $m$ |
| $L$ | load induced by all fragments |
| $U$ | uniform load over all hosts |
| $CL_h$ | current load of host $h$ |
| $SC_h$ | storage capacity of host $h$ |
| $S_t$ | space for storing a triple $t$ |

Table 1: Notation used in Sections 4 and 5

The global query graph $G(QL)$ for query load $QL$ has the anonymized triple patterns of all queries $q \in QL$ as nodes and an undirected edge $\{p, r\}$ iff there is at least one query $q \in QL$ that contains patterns $p'$ and $r'$ with $\omega(p') = p$, $\omega(r') = r$, and $p'$ and $r'$ share a common variable, i.e., they join in $q$; we say that $q$ is a witness for the edge $\{p, r\}$. The weight of such an edge $\{p, r\}$ is the number of its witnesses in $QL$. We further assign a frequency $f(p)$ to each node $p$ of $G(QL)$ which corresponds to the number of queries from $QL$ that contain $p$.

As an example, consider the following query load $QL$, focusing on triple patterns and filter expressions only:

- `?s rdf:type db:city . ?s db:located db:Germany . ?s db:name ?n .` (2x)

- `?s rdf:type db:city . ?s db:located db:USA . ?s db:population ?p .`

- `?s rdf:type db:company . ?s db:located db:Germany .`

- `?s db:name ?c . ?s db:revenue ?r . FILTER(?r≥ $10^9$$)`

- `?s db:name "Apple" . ?s db:revenue ?r .` (10x)

Assuming a normalization threshold of 2 ($\Theta = 2$), we replace constants `db:USA` and `db:company` by a variable but keep `db:Germany`, `db:city`, and `"Apple"`. The corresponding global query graph $G(QL)$ is depicted in Figure 2.

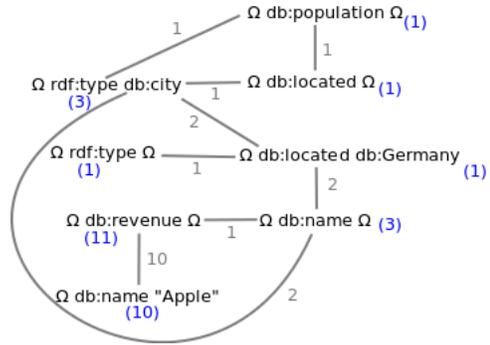

Figure 2: Example Global Query Graph $G(QL)$

### 4.2 Fragmentation

Given a set of triples $T$ (content of the triple store) and a query load $QL$, the second step is to determine appropriate partitions. Similar to horizontal fragmentation of relations [4, 23], we extract a set of simple predicates from $QL$. A *simple predicate* is a constraint on a triple component (subject, property, or object) and either (i) compares a triple component to a constant or URI, i.e., predicates of the form *a op const*, where $a \in \{subj, prop, obj\}$, $op \in \{<, \leq, =, \geq, >\}$, and *const* is a constant literal or URI or (ii) evaluates a function with a boolean result such as $isIRI()$ and value-based comparisons to constants, using again one of $\{subj, prop, obj\}$ to indicate on which component of a triple the function is applied. We do not consider more complex expressions such as $subj = obj$.

To build the set $S(QL)$ of simple predicates for query load $QL$, we consider the anonymized triple patterns as well as the filter expressions. For each $t \in \Phi(QL)$, each position that does not contain $\Omega$ creates a simple predicate. The set of simple predicates derived from the example query load introduced above consists of: `prop=rdf:type`, `prop=db:located`,

`prop=db:population`, `prop=db:revenue`, `prop=db:name`, `obj=db:city`, `obj=db:Germany`, `obj="Apple"`, and `obj`$\geq 10^9$\$.

Based on the set of simple predicates $S(QL)$, we generate the set $M'$ of minterms, i.e., all conjunctive combinations of simple predicates in their positive or negated form, or formally [4]

$$M' := \{ \bigwedge_{p \in S(QL)} p^* \} \qquad (1)$$

where $p^*$ is either the simple predicate $p$ itself or its negation $\neg p$. Each minterm defines a partition of the data set $T$ and the set of all minterms $M'$ partitions all triples in $T$ into $|M'|$ non-overlapping partitions.

The number of minterms is exponential in $|S(QL)|$ and is therefore reduced in two ways: (1) we reduce $S(QL)$ to a subset that is complete (i.e., contained predicates generate fragments that have triples with uniform access patterns) and minimal (i.e., all its predicates influence fragmentation and a fragmentation caused by a predicate is accessed differently by the queries in the query load). We apply the COM_MIN algorithm [23], which, given a set of simple predicates as input, iteratively constructs a set of simple predicates ensuring minimality and completeness in each step. Furthermore, (2) we eliminate all minterms containing contradicting conditions, i.e., contradicting conditions on the same triple component (subject, property, or object) applying knowledge about the semantics of functions that may be called, e.g., $isLiteral(obj)$ and $isIRI(obj)$ will not return *true* for the same $obj$.

This process leads to a smaller set $M$ of minterms with fewer predicates. Each such minterm $m \in M$ defines a fragment $T_m \subseteq T$ of the set of triples, consisting of all triples that satisfy the minterm predicate $m$. These fragments form a complete fragmentation of $T$ since any triple is assigned to at least one fragment by construction of the minterms. They also form a disjoint fragmentation of $T$ since no triple is assigned to multiple fragments, again by construction of the minterms. For notational simplicity, we will identify fragment $T_m$ by its corresponding minterm $m$.

Applying these considerations to our example, we remove redundant predicates and contradicting minterms. Let us assume that the example data set provides information about 3000 cities and 2000 companies all with revenues greater than $10^9$\$ and it does not provide location information for all resources. Then, the predicate `obj`$\geq 10^9$\$ is redundant and therefore removed because the minterms containing this predicate in its positive form are not affected by its presence or absence (the predicate is always true) and minterms containing the predicate in its negative form generate empty fragments – minimality of $M$. After having removed redundant predicates and minterms with contradicting constraints, e.g., `prop=db:located` and `prop=db:population`, we obtain the set of minterms $M$, or fragment definitions respectively, listed in Table 2 – along with access frequency, size, and expected load. In every row, the symbol $\zeta$ denotes all not explicitly mentioned predicates in their negated form, e.g., $\zeta := \neg p_1 \wedge \neg p_2 \cdots \wedge \neg p_k$.

## COM_MIN algorithm

The key consideration that the COM_MIN algorithm relies on is that in each step, it evaluates another simple predicate and compares the obtained fragmentation with it to the fragmentation without it. For this purpose, the algorithm builds all possible minterms and determines access frequencies and query load for each fragment defined by a minterm. By comparing the signatures of the two fragmentations, i.e., the expected load for each fragment $L(m)$ (determined using the query load, the access frequency $f(m)$ and the fragment size $s(m)$: $L(m) := f(m) \cdot s(m)$), the algorithm can efficiently determine if adding the predicate is an improvement (signature changed) or not.

For each fragment $m$, we need to compute its access frequency $f(m)$ and its size $s(m)$. Given a minterm $m \in M$ and a triple pattern $p \in \Phi(QL)$, we say that $m$ *overlaps with* $p$ if there is at least one triple that matches both $m$ and $p$. The access frequency $f(m)$ of $m$ is then the sum of the frequencies of triple patterns from $\Phi(QL)$ that overlap with $m$.

We could evaluate the size $s(m)$ of $m$ by evaluating $m$ over all triples in $T$, but that would often be too expensive. Instead, we consider only a sample large enough to index with a centralized RDF store (we use the open-source system RDF-3X [22]), and evaluate the query on the sample or, even more efficient, use engine statistics to estimate the number of results; the latter is usually only possible for equality predicates.

### 4.3 Fragment Allocation

Once fragments have been defined, they need to be allocated to the $n$ hosts in our cluster. On the one hand, we want to assign fragments that are used together in a (part of a) query to the same host (goal 1) so that the query can be executed locally on that host without transferring intermediate results between hosts. On the other hand, we want to balance the load over all nodes in the cluster (goal 2).

To define the load $L(m)$ attributed to a fragment $m$, we consider the number of queries $f(m)$ for which that fragment is accessed, and assume that all $s(m)$ triples of $m$ are accessed for each such query. Consequently, we get $L(m) := f(m) \cdot s(m)$. Given a fragmentation $M$, the total load $L$ on our system is therefore

$$L := \sum_{m \in M} L(m) \qquad (2)$$

When the load is uniformly balanced over all $n$ hosts in our cluster, each node is assigned a load of

$$U := \frac{L}{n} \qquad (3)$$

Given fragmentation $M$ and query load $QL$, we now define the global fragment query graph $G(QL, M)$ which has the fragments as nodes and an undirected edge $\{a, b\}$ whenever there are triple patterns $p, r \in \Phi(QL)$ such that $a$ overlaps with $p$, $b$ overlaps with $r$, and there is an edge $\{p, r\}$ in $G(QL)$; the weight $w(a, b)$ of the edge $\{a, b\}$ is the sum of the weight of all such edges in $G(QL)$ (i.e., the number of queries that contain a join of triple patterns that overlap with $a$ and $b$, respectively). Figure 3 shows the global fragment query graph $G(QL, M)$ for our running example.

We denote the set of fragments assigned to host $h$ as $F_h$, and the current load of host $h$ as $CL_h := \sum_{m \in F_h} L(m)$. Initially, no fragments are assigned to any host, thus $F_h = \emptyset$ and $CL_h = 0$ for all hosts $h$.

We can now define the benefit of allocating a currently

| | Minterm | Freq. | Size | Load |
|---|---|---|---|---|
| 1 | prop=db:revenue ∧ ζ | 11 | 2000 | 22000 |
| 2 | prop=db:name ∧ ζ | 2 | 4499 | 9998 |
| 3 | prop=rdf:type ∧ obj=db:city ∧ ζ | 3 | 3000 | 9000 |
| 4 | prop=db:population ∧ ζ | 1 | 3000 | 3000 |
| 5 | prop=rdf:type ∧ ζ | 1 | 2000 | 2000 |
| 6 | prop=db:located ∧ ζ | 1 | 1700 | 1700 |
| 7 | prop=db:located ∧ obj=db:Germany ∧ ζ | 3 | 300 | 900 |
| 8 | prop=db:name ∧ obj=Apple ∧ ζ | 10 | 1 | 10 |
| 9 | ζ (Remainder fragment) | 0 | 3000 | 0 |

Table 2: List of Minterms $M$ (fragments) Obtained for Our Example.

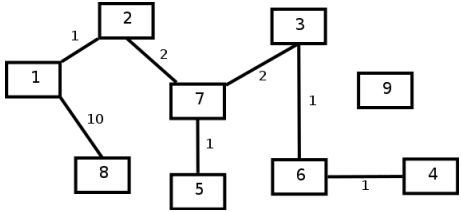

Figure 3: Global Fragment Query Graph.

unallocated fragment $m$ to host $h$ as

$$benefit(m,h) := \frac{2 \cdot U}{U + CL_h} \cdot \sum_{m' \in F_h} \left( w(m,m') + 1 \right) \quad (4)$$

The first part of this definition results in a higher benefit for allocating a fragment to a host which is currently underloaded or in a reduced benefit for already overloaded hosts (goal 2). The second part of the definition results in a high benefit for allocating fragments to a host where many fragments are already allocated that join frequently with this fragment (goal 1).

In addition, we have to consider the space constraint, i.e., a fragment $m$ with size $s(m)$ can only be assigned to a host $h$ if its storage capacity $SC_h$ is not exceeded. Denoting the average storage space required for a triple as $S_t$, we obtain the constraint:

$$SC_h \geq s(m) \cdot S_t + \sum_{m' \in F_h} (s(m') \cdot S_t) \quad (5)$$

For allocating fragments to hosts, we use the following greedy algorithm:

- We allocate fragments in descending order by their load $L(m)$.

- For each fragment, we calculate the benefits of allocating it to every host and assign it to the most beneficial host for which the space constraint is not violated.

Assume we need to allocate the fragments we generated for our running example to three hosts ($n = 3$). We start with allocating fragment 1 (highest load, Table 2) to a random host (host 1). For Fragment 2 we determine the benefits of assigning it to host 1 (benefit: 0.85), host 2 (benefit: 2), and host 3 (benefit: 2) – we assign it to host 2. After having assigned all fragments to hosts, we obtain the following allocation: host 1 is assigned fragments 1 and 8, host 2 is assigned fragment 2, and host 3 is assigned fragments 3, 4, 5, 6, and 7. Fragment 9 is receiving special treatment as explained below.

### Avoiding Imbalance

The construction of minterms based on predicates occurring in queries (Section 4.2) produces one minterm that contains all predicates in their negated form, i.e., the corresponding fragment contains all triples that do not match any of the simple predicates contained in the queries of the training data set. In many cases, this "remainder" fragment is much bigger than all the other fragments and would lead to a highly imbalanced usage of storage space at the sources. Thus, in case it is significantly bigger than the other fragments, we need to split it up into smaller fragments and assign it to other partitions.

To integrate this consideration into our implementation, all fragments but the remainder fragment are assigned as described above. Afterwards, we apply hash partitioning to assign the triples of the remainder fragment to partitions. The reason is that, in case of all the other fragments the global query optimizer can efficiently decide on the relevance of a fragment by comparing the fragment definition to the predicates contained in the query. In any other case, the optimizer has to use statistics or query all available fragments. Thus, the decision whether to ask two or all fragments of roughly the same sizes has only little influence on response time because queries are processed in parallel at different hosts whereas querying a much bigger fragment would increase response time.

### Handling Updates

PARTOUT can efficiently handle updates: when a new triple is inserted, the coordinator uses the fragment definitions to determine the responsible partition and sends the triple to the corresponding host. The procedure for deletions uses the same principle. If a modification of a triple does not affect allocation, the corresponding host handles it locally. Otherwise, the coordinator converts the update into a deletion followed by an insertion and notifies the corresponding hosts. We expect that inserts and deletes are usually submitted in batches and can therefore be handled efficiently. If a host is heavily overloaded as a result of many newly inserted triples allocated to it, some of its assigned fragments might be relocated to other hosts.

### Bootstrapping

The process of distributing the triples in $T$ to the hosts is run after the allocation is finished. The central coordinator first scans $T$ once to build a global dictionary of strings (i.e., URIs and literals) and a unique mapping of these strings to integer ids, and distributes this mapping to all hosts. If $T$ can be accessed from every host, each host then independently scans $T$ and inserts the triples matching its assigned

fragments into its local triple store. If $T$ can be accessed only at the central coordinator, the coordinator scans $T$, collects triples to be stored at the hosts, and sends them to their target host. Although our current implementation uses a central coordinator for bootstrapping, this process can be easily parallelized. The global dictionary, for example, can be built in parallel by splitting $T$ into multiple temporary partitions (e.g., using a hash function), creating dictionaries for the partitions in parallel, and merging these dictionaries afterwards.

## 4.4 Determining the Optimal Number of Hosts

The allocation algorithm introduced above assumes a fixed number of hosts $n$ as input. To find the optimum trade-off between communication costs during query processing and load balancing, the configuration (number of hosts $n$) resulting in the minimum response time for the given query load should be chosen. Thus, to find the optimal number of hosts, we run the allocation algorithm for different values of $n$ and choose that value for $n$ that produces the minimum estimated response time for the given query load – using the query optimizer and cost model introduced in Section 5 for estimation.

## 5. DISTRIBUTED QUERY PROCESSING

Having set up a distributed system as explained above, we also need a query optimizer that generates efficient query execution plans. In the following, we first introduce the general setup of software components and statistics, and then provide details on how our optimizer works and how a query is evaluated.

## 5.1 System Setup and Statistics

As Figure 1 illustrates, distributed query processing in PARTOUT involves two software components: a coordinator and multiple slaves – both based on RDF-3X [21, 22]. Queries are issued at the coordinator, which is responsible for generating a suitable query plan for distributed query execution. The data is located at the slaves, each managing a set of fragments of the original database (referred to as partition). The slaves execute the parts of the query assigned to them over their local data and send the results to the coordinator, which will finally hold the query result.

### Coordinator

The coordinator does not have direct access to the actual data but instead uses a global statistics file generated at partitioning time containing the following information:

- *Fragment definitions, their sizes, and host mappings.* Statistics contain information about how the fragments were created, their cardinality, and which hosts they were assigned to.

- *String dictionary* with the unique string-to-integer mapping used in all hosts, i.e., not the original strings are stored but the ids they were mapped to. The dictionary is built during bootstrapping (Section 4.3).

- *Statistics.* The coordinator estimates all necessary statistics based on a representative sample, e.g., the same one that was already used in Section 4.2 to estimate sizes of potential fragments. Alternatively, if the coordinator has enough free storage space, it can also retrieve detailed statistics directly from the slaves using caching or aggregation to limit the amount of used space. The statistics we work with to estimate costs for operators, such as index scans, are the number of entries and the number of pages necessary to store the data.

To avoid a single point of failure and bottlenecks, this information can be replicated among several hosts.

### The Slaves

Slaves are lightweight server processes running at different machines. They are listening for execution requests coming either from the coordinator or from other slaves and execute them over their local partition. With all necessary statistics being available at the coordinator, slaves do not need a local query optimizer.

## 5.2 Global Query Optimization

Optimizing for response time, our goal is to find query execution plans that minimize the time that elapses from the initiation of a query to its completion. PARTOUT's global query optimization algorithm avoids exhaustive search as estimating costs for all possible plans would be too expensive in a distributed environment because of the additional options for query execution. We apply a two-step approach starting with a plan optimized with respect to cardinalities and selectivities and then apply heuristics to obtain an efficient plan for the distributed setup. As the RDF-3X optimizer is known to produce efficient query execution plans, we have the coordinator create such a plan using the available statistics and use it as input for the second step. In the second step, we rely on a cost model that helps us identify good plans for distributed query execution.

### Initial Query Plan

RDF-3X, and therefore each PARTOUT slave, maintains local indexes for all possible orders and combinations of the triple components (and for aggregations), which enable efficient local data access. Hence, leaf-level operators in query plans are index scans corresponding to triple patterns in the query. For instance, the triple pattern (`?s,rdf:type,db:city`) results in a scan of the $POS$ index, retrieving matching subjects for given property and object values in increasing id order. This is exploited for join operators which are implemented by efficient merge joins whenever the two inputs are ordered by the join attribute; if that is not the case, the inputs need to be sorted or a hash join is applied. For more details about the RDF-3X optimizer, which uses dynamic programming, we refer the reader to the respective literature [21, 22].

### Distributed Cost Model

In general, the initial query execution plans produced by the RDF-3X optimizer use pipelining whenever possible, i.e., most operators will report results to the next operator in the hierarchy before the result has been computed completely. The only exception to this rule are pipeline breakers, such as the sort operator, that need to read the complete input before producing the first result. To minimize communication costs in the distributed setup, we do not transfer each tuple individually to another host but in batches (pages) of 1024 results.

Finding the best query execution plan requires a cost function $c(plan)$ that determines the execution costs for a plan $plan$. As defined in Equation 6, the costs $c(plan)$ of a plan are defined as the costs $c(RootOp)$ of its root operator.

$$c(plan) = c(RootOp) \qquad (6)$$

The costs of an operator $op$ consist of its execution costs $xc(op)$ (the estimated response time to evaluate it, which can be estimated by the RDF-3X optimizer), the costs of its child operators (because of exploiting parallelism in combination with response time, we only need to consider the maximum of all children), and possibly the costs $tc(x, op)$ to transfer the results between hosts:

$$c(op) = xc(op) + \max_{x \in children(op)} (tc(x, op) + c(x)) \qquad (7)$$

We can estimate the costs to transfer results from operator $x$ to operator $op$ as:

$$tc(x, op) = \begin{cases} t_{page} \cdot \left\lceil \frac{excard(x)}{1024} \right\rceil & ; hh(x) \neq hh(op) \\ 0 & ; \text{otherwise} \end{cases} \qquad (8)$$

$hh(op)$ denotes the home host of an operator and $excard(op)$ is the expected output cardinality of operator $op$ as estimated by the RDF-3X optimizer. We assume that communication costs for one page of data are symmetric, constant, and equal to $t_{page}$, which has to be empirically measured for a specific setup. Still, $t_{page}$ can easily be replaced by more complex models and hence be adapted to environments with heterogeneous network topologies and distances between hosts.

## Query Planning

The leaf nodes of the initial query plan represent index scans on a global database corresponding to triple patterns in the query. In our setup, however, the triples matching these triple patterns are spread across the hosts. Thus, we first need to identify the hosts that are relevant to the leaf node scans.

Using the coordinator's statistics, more specifically fragment definitions and host mappings, we compare the triple pattern to the fragments' minterm definitions and thus identify if a host's partition contains a fragment with relevant data. If the triple pattern overlaps with the minterm of the remainder fragment, all hosts might be relevant in dependence on its treatment during allocation (Section 4.3). The leaf nodes in the initial plan are replaced accordingly: if there is only a single relevant host for a leaf operator, the scan for the triple pattern is exclusively executed at that host. If there is more than one relevant host, the scan is executed at each host and their results are combined through a chain of binary merge-union (BMU) operators preserving sort order.

In the obtained plan, leaf operators in the query plan are annotated with their execution site (*home host*), i.e., the hosts holding the data the operator refers to. In the next step, the optimizer has to assign home hosts to inner operators as well – with $hh(op)$ denoting the home host of operator $op$. Moreover, the optimizer considers to replace the chains of BMU operators with distributed joins. Thus, the optimizer recursively traverses the operator tree, assigns hosts to the inner nodes, applies transformations, and uses the cost model to find an efficient query execution plan.

In a bottom up fashion, starting at the leaf nodes, the home host for an operator on a higher level is selected among the set of home hosts assigned to its children and the candidate home host for the root operator – the optimizer considers each available host as a candidate for the root operator, which generates good candidate execution plans that would not be found using a purely greedy approach based on the hosts of child operators only. For chains of BMU operators, the optimizer does not consider each BMU operator in separate but assigns the same home host to all BMU operators and the parent operator (e.g., a join) of the topmost BMU operator.

The optimizer also considers transformations to the query plan motivated by heuristics if applicable and evaluates the benefit of the transformation using the cost model. One of them is to prefer merge joins over hash joins, another one is to push down selections and projections contained in the query using standard algebraic transformation rules exploiting associativity, commutativity, and distributivity rules. Another transformation tries to improve the plan by introducing distributed joins that exploit pipelining. For example, for a join involving multiple hosts (i.e., a chain of BMU operators), the optimizer (i) identifies all relevant combinations of partial joins between involved hosts, (ii) adds a MergeJoin operator for each combination, and (iii) combines partial results using binary merge union (BMU) operators. Especially for non-selective queries with many intermediate results, this transformation can significantly reduce response time because many tuples can be processed in parallel.

### Example

Figure 4 shows an example query execution plan obtained by applying the algorithm above on the following query:

```
SELECT ?NAME WHERE {
    ?s rdf:type db:city .
    ?s db:located db:Germany .
    ?s db:name ?name .  }
```

According to the statistics, hosts $H1$, $H2$, and $H3$ hold relevant data for the first triple pattern, only $H1$ for the second, and only $H2$ for the third. For local access to the triples at each host, the POS (predicate-object-subject) index is chosen by the RDF-3X optimizer. The optimizer added a chain of BMU operators to scan the data for the first triple pattern and computes the join with the second triple pattern, and therefore with the data provided by host $H1$, using a chain of efficient MergeJoins. According to the cost model, the optimizer decided to have the MergeJoins executed at host $H1$, which uses local pipelining to execute the assigned parts of the query plan efficiently in parallel.

Note that, although in this example most of the computation is done using pipelining at the same host, in general bigger parts of the query are executed in parallel at different hosts. Furthermore, load balancing can be improved by extending the cost model to consider the load generated by currently running queries.

### 5.3 Query Execution

For query execution, the optimized query plan is extended with operators for exchanging data between hosts (*Remote Sender* and *Remote Fetcher* operators). Thus, whenever the home host of a child node differs from the one of its parent node, the two operators are inserted. Before query execution can start, subqueries are identified and sent to the corresponding home hosts for execution.

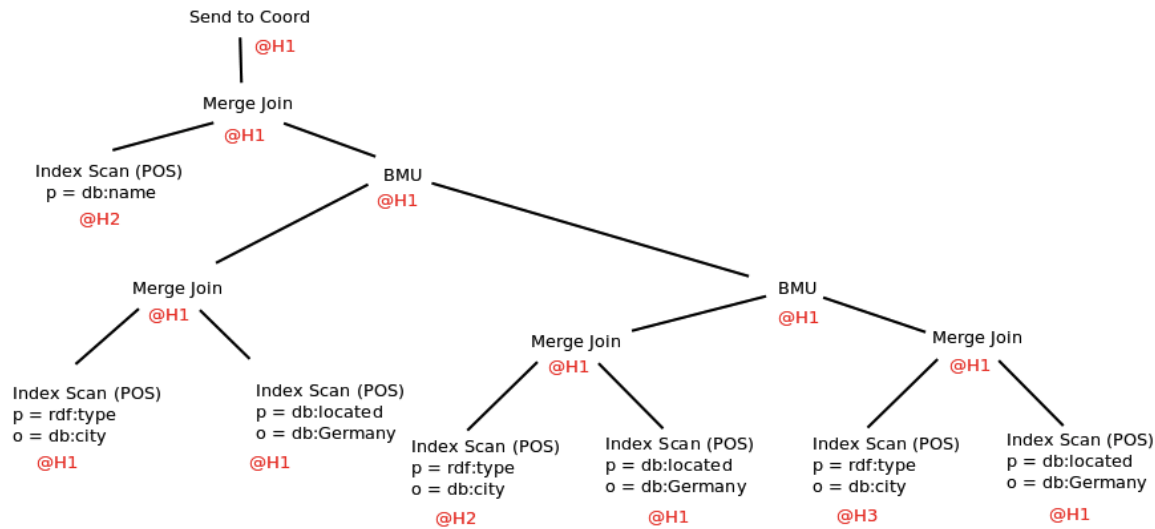

Figure 4: Example query execution plan after optimization

Query execution at each host starts as soon as inputs to an operator are available, e.g., local index scans, intermediate results of local operators, or results retrieved via *Remote Fetcher* operators. Results of root nodes are forwarded to other hosts according to the corresponding *Result Sender* operator when either one page of results is filled (pipelining) or all results have been generated.

## 6. EVALUATION

To evaluate PARTOUT's partitioning technique, we compared it against a variant (fragmentation by property) used by state-of-the-art approaches that defines partitions using predicates contained in the query load and that does not consider the co-occurrence of predicates in the queries for fragment allocation. We also measured performance using the centralized RDF-3X version that PARTOUT is based upon.

Moreover, we compared PARTOUT's partitioning technique to the 2-hop directed partitioning proposed by Huang et al. [14], i.e., after having applied a graph-partitioning algorithm on the complete data set (RDF graph), triples at the borders of the obtained partitions are replicated. Assume, for instance, the node db:AngelaMerkel is a border node and assigned to partition 1, then all triples with this node as subject are assigned to partition 1, e.g., (db:AngelaMerkel, db:livesIn,db:Berlin). In addition, all triples connected to the border node in a distance of 2 hops (e.g., (db:Berlin, db:locatedIn,db:Germany)) are replicated to partition 1 although their subject nodes and therefore the triples are assigned to other partitions. The replication allows to evaluate queries in one partition although relevant triples are originally assigned to other partitions. If a query cannot be answered completely on a single partition (e.g., the diameter of the query graph pattern exceeds a threshold – a so-called non-PWOC query), Huang et al. propose the use of MapReduce, which results in a start-up overhead of about 20 seconds [14]. As our main interest concerns the proposed partitioning technique, our implementation avoids the expensive use of MapReduce; instead, we split up a non-PWOC query into subqueries that each can be handled by a single partition, and use a pipelined query plan involving hash joins and union operators. Therefore, execution times for non-PWOC queries are much lower than they would have been with the original system because we save the substantial overhead that comes along with MapReduce. In the following, we will refer to this implementation as HAR+.

## Setup

All experiments were implemented on top of RDF-3X version $0.3.6^{12}$ using a cluster of machines each with an Intel Xeon E5430 processor and 32 GB of RAM running Debian GNU/Linux 6.0 for 64 bits. The machines ran instances of PARTOUT's lightweight slave component each managing one data partition. An additional machine ran the coordinator component; an Intel Xeon E5530 processor, 32 GB of RAM running Debian GNU/Linux 5.0.9 for 64 bits.

## Datasets

Sp2Bench is one of the datasets of the FedBench benchmark[13] and, as part of FedBench, includes 10M RDF triples of synthetic bibliographic information. The second dataset originates from the Billion Triple Challenge (BTC) 2008[14] and was already used in [21] to evaluate RDF-3X. It contains more than 500M triples originating from different datasets that are part of the Linked Open Data cloud. Note that we could not use the Lehigh University Benchmark (LUBM) [10] used by Huang et al. [14] because most of its queries require inferencing, which is not supported by RDF-3X and therefore lead to empty results.

### 6.1 Billion Triple Challenge (BTC)

As no standard benchmarks are available for this use case, we generated 30 random queries from the complete data set by randomly picking properties from the data, using the property as a seed to generate a triple pattern, and extending the triple pattern according to the available data to SPARQL queries (limiting the max. number of triple

---
[12] http://code.google.com/p/rdf3x/
[13] http://code.google.com/p/fbench/
[14] http://challenge.semanticweb.org/

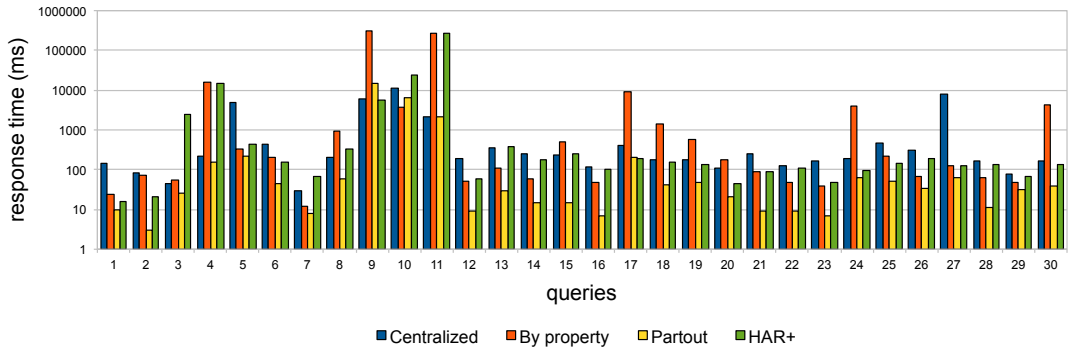
(a) 3 hosts

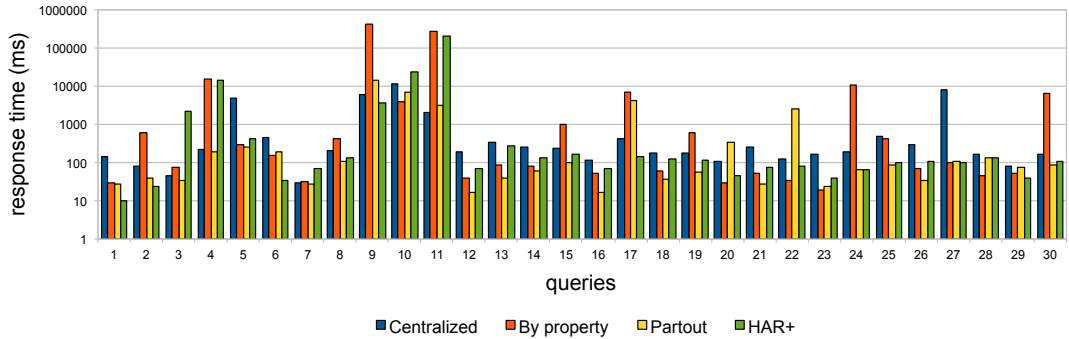
(b) 5 hosts

Figure 5: Response Times for the Billion Triple Challenge (BTC) Data Set

patterns for a star-shaped query to 6 and for a path-shaped query to 4). For presentation, we have ordered the queries according to their main characteristics, i.e., queries 1 and 2 consist of a single triple pattern, queries 3 to 6 are path-shaped queries, queries 7 to 9 are double-star-shaped path queries (two connected star query patterns), queries 10 to 30 correspond to star-shaped queries. With respect to the technique proposed by Huang et al. queries 3, 4, 5, 7, 10, 11, 12 correspond to non-PWOC queries.

Having determined the optimal number of hosts for our setup (Section 4.4) to be 3, we present our results for a PARTOUT configuration with 3 and 5 slave hosts in Figure 5. The first observation is that the workload-aware partitioning technique proposed in this paper outperforms by-property-partitioning and therefore state-of-the-art approaches relying on such a technique. The reason for the gain in performance is the consideration of co-occurrences of triple patterns in queries during partitioning and allocation. Ignoring these co-occurrences, fragments corresponding to triple patterns involved in a join are likely to be assigned to different hosts and much data needs to be transferred to answer a query.

For 3 hosts, PARTOUT even outperforms the centralized approach for almost all the queries. The reason is that in PARTOUT, query processing benefits from parallel query processing on smaller partitions, i.e., parts of the query are executed in parallel and evaluating a query on a smaller fragment is faster than on the complete original data set. Furthermore, having optimized the partitioning with respect to a particular query load, PARTOUT can often identify a single partition that can answer the query completely, eliminating the need for costly communication between hosts. In the experiments for 5 hosts, we see the negative influence of communication on some queries, i.e., response time for some of the queries is higher than for 3 hosts because relevant fragments have been assigned to different hosts for the benefit of load balancing. Note that when setting up the system (Section 4.4), PARTOUT would have identified 3 hosts to be the optimum.

Similar observations hold for HAR+. Whereas PARTOUT often identifies a relevant partition, for HAR+ (because of the absence of fragment definitions) all partitions (hosts) have to participate (run the query in parallel) and for non-PWOC queries compute a join across partitions. Furthermore, the necessary removal of duplicates introduced by the replication increases processing time for HAR+, even though we use an additional `isOwned` triple to indicate the original partition of a triple (i.e., the partition of its subject) [14]; however, this results in an additional join during query evaluation.

### 6.2 Sp2Bench

Figure 6(a) reports our results for the FedBench[13] Sp2Bench benchmark with 2 hosts. We used a modified subset of the benchmark queries; we removed components and queries (such as optional, offset, etc.) that RDF-3X does not support. The dataset is rather small in comparison to the BTC dataset – only 10 million RDF triples. In fact, the approach outlined in Section 4.4 would report that 1 host is the optimal configuration. The benchmark queries were designed to generate high load on a system, i.e., expensive queries with joins whose evaluation basically involves all

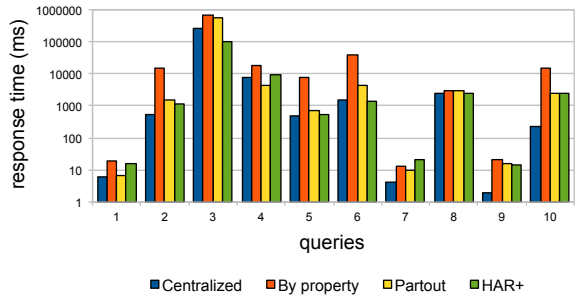
(a) Response Times for Sp2Bench with 2 hosts

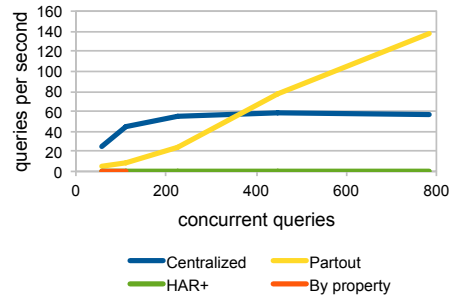
(b) Throughput for the BTC data set with 3 hosts

Figure 6: Response Times for Sp2Bench and Throughput for BTC

the data in the system. The results show that even though evaluating most of these queries involves all the data of all partitions and PARTOUT has the disadvantage of communication costs, PARTOUT's performance is still competitive in comparison to the other approaches, especially the centralized approach.

### 6.3 Throughput

We also ran experiments to compare the approaches with respect to throughput at different levels of concurrency – we issued the set of queries several times in intervals of 1 second. The results for the BTC data set running 3 hosts are depicted in Figure 6(b). With all queries handled by one machine, the centralized approach does not scale that well for this large data set. The number of concurrent queries is also limited for the by-property-partitioning approach because of the high level of necessary communication between hosts – tests with more than 120 concurrent queries already ran into a timeout. Similar problems occur with HAR+; with the need of sending each query to all hosts and removing duplicates, high load is generated for all machines and it is hence difficult to handle a large number of concurrent queries. By sharing the load among 3 hosts, PARTOUT can efficiently handle concurrent queries because for each query that can be answered by a single host, load at all the other nodes is not affected.

### 6.4 Discussion

The evaluation results show that PARTOUT's performance in a distributed setup is competetive to other architectures, even the centralized one. Especially in terms of throughput PARTOUT scales much better than any of the other approaches because PARTOUT does not always require the participation of all hosts when evaluating a query, and each host keeps only a subset of the data.

In our current implementation, we do not consider replication, i.e., each triple is stored only once. In comparison to other approaches this reduces the overall amount of data that has to be stored in the whole system. At the same time, it alleviates the problem of handling duplicates. More importantly, it enables PARTOUT to efficiently handle updates, which is, as discussed in Section 4.3, much more difficult and expensive in other distributed systems using replication.

For query execution, however, PARTOUT would benefit from replication and some queries could be executed even more efficiently. We can extend PARTOUT with replication by running the fragmentation and allocation algorithms as explained in this paper and replicating fragments to hosts acccording to the benefit and the global fragment query graph in a second round. In so doing, PARTOUT would also better scale with the number of hosts and might in extreme cases lead to a system where the data is replicated to all nodes. However, as a consequence of such a replication, updates would be less efficient, query optimization would become more complex, and local query execution at the hosts would be slower because of the larger data sets they have to manage, so throughput could decrease.

As the focus of this paper is partitioning the data according to a query load and showing that an appropriate distributed approach without replication can already compete with existing solutions, we consider replication an orthogonal problem and will consider it in detail in our future work.

### 7. CONCLUSION

In this paper, we have proposed PARTOUT, a distributed system for evaluating queries over large amounts of RDF data. Building upon a state-of-the-art triple store for centralized query execution, this paper focused on higher-level problems that come along with the distribution: data partitioning, allocation, and distributed query processing. In consideration of the query load, we presented algorithms for partitioning the data into fragments and assigning the obtained fragments to the set of available hosts so that fragments that are often needed in combination to evaluate a query are located at the same host. We also presented a query optimizer along with a distributed cost model and several heuristics that lead to efficient query execution plans that also exploit pipelining. Our evaluation results show that the proposed techniques are suitable for the distributed setup and especially speed up queries involving large intermediate results that in combination with a beneficial allocation scheme do not have to be transferred to another host. Moreover, throughput scales much better with the number of queries in comparison to alternative approaches. Our experiments also showed that setting up a distributed system in consideration of replication and without consideration of a query load has advantages in some scenarios, which are, however, very often outweighed by the need to handle duplicates during query processing and the absence of means to identify irrelevant hosts. Nevertheless, in our future work we will try to combine the advantages of query-load aware partitioning and replication in one system trying to avoid the disadvantages we have identified. PARTOUT in its cur-rent state can already efficiently handle updates on the data. In our future work, we plan to consider also changes in the query load, i.e., finding answers to the question how a running system might be adapted efficiently and identify when it is worthwhile to reconfigure.